\documentclass[conference,a4paper]{IEEEtran}
\IEEEoverridecommandlockouts

\usepackage{xcolor}
\usepackage{balance}

\usepackage{cite}
\usepackage{amsmath,amssymb,amsfonts}
\usepackage{graphicx}
\usepackage{textcomp}
\usepackage{acronym}
\usepackage{xcolor}
\usepackage{tikz} 
\usepackage[utf8]{inputenc}
\usepackage{pgfplots} 
\usepackage{pgfgantt}
\usepackage{pdflscape}
\usepackage{changes}
\usepackage{comment}
\usepackage{subfigure}
\usepackage{mathtools,algpseudocode,algorithm,MnSymbol}
\usepackage{geometry}
\usepackage{pgfplots}
  \pgfplotsset{compat=newest}
  \usetikzlibrary{plotmarks}
  \usetikzlibrary{arrows.meta}
  \usepgfplotslibrary{patchplots}
  \usepackage{grffile}
  \usepackage{amsmath}

\pgfplotsset{compat=newest} 
\pgfplotsset{plot coordinates/math parser=false} 

\acrodef{MMSE}{Minimum Mean Squared Error}

\acrodef{MSE}{mean square error}

\acrodef{PSD}{power spectral density}

\acrodef{RMSE}{root mean squared error}
\acrodef{SLR}{statistical linear regression}

\acrodef{IPLF}{iterated posterior linearization filter}

\acrodef{ue}[UE]{user equipment}
\acrodef{bs}[BS]{base station}
\acrodef{va}[VA]{virtual anchor}
\acrodef{sp}[SP]{scattering  point}
\acrodef{fov}[FoV]{field-of-view}   
\acrodef{los}[LOS]{line-of-sight}
\acrodef{nlos}[NLOS]{non-line-of-sight}
\acrodef{PMBM}[PMBM]{Poisson  multi-Bernoulli  mixture}
\acrodef{TPMBM}[TPMBM]{trajectory Poisson  multi-Bernoulli  mixture}
\acrodef{PMB(M)}[PMB(M)]{Poisson  multi-Bernoulli  (mixture)}
\acrodef{PMB}[PMB]{Poisson  multi-Bernoulli}
\acrodef{rfs}[RFS]{random finite set}
\acrodef{PPP}[PPP]{Poisson point process}
\acrodef{MBM}[MBM]{multi-Bernoulli  mixture}
\acrodef{ekf}[EKF]{extended Kalman filter}
\acrodef{PDF}[PDF]{probability density function}

\acrodef{ckf}[CKF]{cubature Kalman filter}
\acrodef{rbp}[RBP]{Rao-Blackwellized particle}
\acrodef{gospa}[GOSPA]{generalized optimal subpattern assignment}
\acrodef{slam}[SLAM]{simultaneous localization and mapping}

\acrodef{TOA}[TOA]{time of arrival}
\acrodef{AOA}[AOA]{angles of arrival}
\acrodef{AOD}[AOD]{angles of departure}

\acrodef{MIMO}{multiple-input multiple-output}
\acrodef{OFDM}{orthogonal frequency-division multiplexing}

\acrodef{URA}[URA]{uniform rectangular array}

\acrodef{DA}{data association}

\acrodef{ISAC}[ISAC]{integrated sensing and communications}

\acrodef{DISAC}[DISAC]{distributed integrated sensing and communications}

\acrodef{RMS}[RMS]{root mean squared}

\acrodef{MTT}[MTT]{multi-target tracking} 
\allowdisplaybreaks

\begin{document}

\bibliographystyle{IEEEtran}
\bstctlcite{IEEEexample:BSTcontrol}

\title{Target Handover in Distributed Integrated Sensing and Communication}

\author{\IEEEauthorblockN{
Yu Ge\IEEEauthorrefmark{1},  Ossi Kaltiokallio\IEEEauthorrefmark{2}, Hui Chen\IEEEauthorrefmark{1}, Jukka Talvitie\IEEEauthorrefmark{2}, Yuxuan Xia\IEEEauthorrefmark{3},  Giyyarpuram Madhusudan\IEEEauthorrefmark{4}, \\ Guillaume Larue\IEEEauthorrefmark{4},
Lennart Svensson\IEEEauthorrefmark{1}, Mikko Valkama\IEEEauthorrefmark{2},      
Henk Wymeersch\IEEEauthorrefmark{1} 
}                                     
\IEEEauthorblockA{\IEEEauthorrefmark{1}
Department of Electrical Engineering, Chalmers University of Technology, Gothenburg, Sweden,} 
\IEEEauthorblockA{\IEEEauthorrefmark{2}
Unit of Electrical Engineering, Tampere University, Tampere, Finland,}
\IEEEauthorblockA{\IEEEauthorrefmark{3}
Department of Automation, Shanghai Jiaotong University, Shanghai, China,\IEEEauthorrefmark{4}
Orange Research, Meylan, France}
\thanks{This work has been supported by the SNS JU project 6G-DISAC under the EU's Horizon Europe research and innovation Program under Grant Agreement No 101139130.}
}

\maketitle

\begin{abstract}
The concept of 6G \ac{DISAC} builds upon the functionality of \ac{ISAC} by integrating distributed architectures, significantly enhancing both sensing and communication coverage and performance. In 6G \ac{DISAC} systems, tracking target trajectories requires \acp{bs} to hand over their tracked targets to neighboring BSs. Determining what information to share, where, how, and when is critical to effective handover. This paper addresses the target handover challenge in \ac{DISAC} systems and introduces a method enabling \acp{bs} to share essential target trajectory information at appropriate time steps, facilitating seamless handovers to other BSs. The target tracking problem is tackled using the standard \ac{TPMBM} filter, enhanced with the proposed handover algorithm. Simulation results confirm the effectiveness of the implemented tracking solution.

\end{abstract}

\vskip0.5\baselineskip

\acresetall

\vspace{-5mm}

\section{Introduction}
\Acf{ISAC} has attracted 
attention in the advancement of 5G mobile radio systems and is expected to play a pivotal role in shaping 6G  networks \cite{An_ISAC_Survey2022,liu2022integrated}. However, the current \ac{ISAC} framework often overlooks critical aspects, such as enabling large-scale deployments that can monitor numerous connected user devices and passive objects over wide areas and extended periods \cite{strinati2024distributed,strinati2024towards,10289611}. To address these limitations, the concept of \ac{DISAC} has emerged, which expands \ac{ISAC} functionality through distributed architectures, enhancing its capabilities in decentralized systems. In addition, \ac{DISAC}  is set to revolutionize various sectors, including transportation, 
and automation \cite{strinati2024distributed,strinati2024towards}.

\begin{figure}
\centerline{\includegraphics[width=0.92\linewidth]{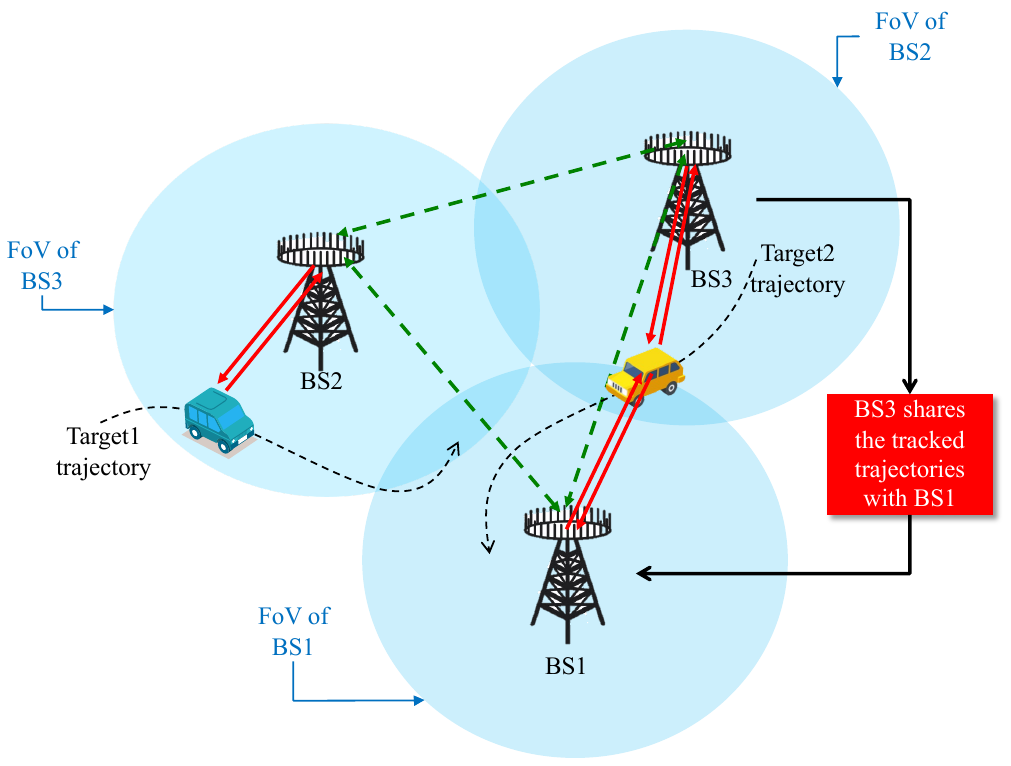}}
\vspace{-2mm}
\caption{An example of the target handover problem in a monostatic DISAC system. Targets move over time, and each \ac{bs} can only observe those within its own \ac{fov}. Each BS independently processes its local observations to track the targets, then shares the tracked trajectories with neighboring BSs (i.e., with overlapping or adjacent \ac{fov}).}
\label{fig:overview}

\end{figure}

In a monostatic DISAC system, several  \acp{bs} transmit orthogonal signals that bounce off the environment and are received by the same BS, ensuring full knowledge of both the transmitted signals and the clock \cite{sturm2011waveform,ge2022mmwave}. Due to the limited transmission power and hardware limitations, each BS can only track targets within its \ac{fov}. As a result, without information sharing between BSs, each BS can only track  target trajectories for a limited time and only within its own \ac{fov}. To enable more comprehensive tracking without extensive data sharing, BSs must \textit{hand over targets} to neighboring \acp{bs} when necessary. This allows the continuation of target trajectories across \acp{bs} while minimizing communication between them. This setup is depicted in Fig.~\ref{fig:overview}. Deciding what information to share, as well as when, where, and how to share it, is crucial for efficient handover.

Related works can be categorized into those addressing the  \ac{MTT} sensing problem systems and those focusing on target handover between different BSs or sensors. The sensing challenge has been explored extensively and explored through various methods in the \ac{MTT} literature \cite{yassin2018mosaic,Erik_BPSLAM_TWC2019,ge2022computationally,ge2022mmwave,ge2023integrated}. Among these, \ac{rfs}-based methods \cite{ge2022computationally,ge2022mmwave,ge2023integrated} can effectively manage key challenges such as unknown \acp{DA}, an uncertain number of targets, unknown target states, missed detections, and false alarms, and can also generate entire target trajectories without increasing the computational complexity \cite{svensson2014target,garcia2019multiple}. 
However, these methods either rely on data from a single source or use a fusion center to combine sensing results from multiple sources, without incorporating any target handover mechanism. The concept of \textit{target handover} has been addressed in various forms \cite{kim2019optimal,yazdi2014study,sonmez2020handover}, primarily in the context of networked radar systems. In \cite{kim2019optimal} a binary linear programming approach is used to schedule radars in order to maximize the tracking duration of targets. In \cite{yazdi2014study}, the concept of handover refers to the association between moving sensors and static coordinators, and is thus  similar to \textit{\ac{ue} handover}  in communication systems \cite{shayea2020key,sonmez2020handover}. Hence, the problem of target handover  remains largely unexplored.

In this paper, we explore the target handover problem in the \ac{DISAC} system, enabling each BS to track complete target trajectories within its \ac{fov} from origin until they exit the \ac{fov}, by sharing necessary trajectory information among BSs. The main contributions of this paper are as follows: \textit{(i)} We propose an RFS-based target handover algorithm that facilitates trajectory sharing between neighboring BSs, ensuring smooth trajectory tracking as the targets enter their \acp{fov}, along with detailed execution steps; \textit{(ii) }We extend the standard \acf{TPMBM} filter at each BS by incorporating the proposed target handover mechanism; \textit{(iii)} We validate the effectiveness of the proposed target handover mechanism through simulations in a mmWave radio network, demonstrating that it allows BSs to track targets' trajectories from their origin, while also enhancing sensing performance when targets transition into  new \acp{fov}.

\subsubsection*{Notations}
Scalars (e.g., $x$) are denoted in italic, vectors (e.g., $\boldsymbol{x}$) in bold lower-case letters, matrices (e.g., $\boldsymbol{X}$) in bold capital letters, sets  (e.g., $\mathcal{X}$) in calligraphic. Transpose is denoted by $(\cdot)^{\top}$, the Hermitian transpose is denoted by $(\cdot)^{\mathsf{H}}$, the union of mutually disjoint sets is denoted by $\uplus$, a Gaussian density with mean $\boldsymbol{u}$ and covariance $\boldsymbol{\Sigma}$, evaluated in value $\boldsymbol{x}$, is denoted by $\mathcal{N}(\boldsymbol{x};\boldsymbol{u},\boldsymbol{\Sigma})$, and $d_{\boldsymbol{x}}=\text{dim}(\boldsymbol{x})$.

\vspace{-4mm}
\section{System Model}
\vspace{-1mm}
We consider a scenario with multiple \acp{bs} and multiple targets, as depicted in Fig.~\ref{fig:overview}. Each \ac{bs} has limited \ac{fov}, and can only see the targets in its \ac{fov}. The \acp{fov} of two neighboring \acp{bs} can overlap. Each individual \ac{bs} performs monostatic sensing to track the moving targets in its \ac{fov}, and it can share the information on target tracks to nearby \acp{bs}. In this section, the state models, the received signal model, and the measurement model are introduced. 

\vspace{-3mm}
\subsection{State Models} \label{Sec:statemodels}
\vspace{-1mm}
Each \ac{bs} is deployed with a \ac{URA}. The state of the $p$-th \ac{bs} contains a 3D location $\boldsymbol{s}_{\mathrm{BS}}^{p}$, representing the position of the center of the \ac{URA}, and its 3D orientation $\boldsymbol{\psi}_{\mathrm{BS}}^{p}$. The number of \acp{bs} and all \ac{bs}
states are known. While multiple targets in the considered environment are moving over time, neither the number of targets nor the target states are known to the \acp{bs}. We denote $\boldsymbol{x}_{k}^{i}=[(\boldsymbol{s}_{k}^{i})^{\top},(\dot{\boldsymbol{s}}_{k}^{i})^{\top}]^{\top}$ as the dynamic state of the $i$-th target at time step $k$, containing its 3D position $\boldsymbol{s}_{k}^{i}$ and its 3D velocity $\dot{\boldsymbol{s}}_{k}^{i}$. We assume the transition density of each target can be described by
\begin{equation}
f(\boldsymbol{x}^{i}_{k+1} | \boldsymbol{x}^{i}_{k}) = {\cal N}(\boldsymbol{x}^{i}_{k+1} ; \boldsymbol{v}(\boldsymbol{x}^{i}_{k}),\boldsymbol{Q}_{k+1}), \label{dynamicmodel}
\end{equation}
where $\boldsymbol{v}(\cdot)$ represents the known transition function and $\boldsymbol{Q}_{k+1}$ is the known process noise covariance.

\subsection{Signal Models}
Every time step, each \ac{bs} sends out downlink \ac{OFDM} pilot signals, which are altered by the objects in the environment and then possibly received by the \ac{bs}. At the $p$-th BS side, the received signal at the subcarrier $\kappa$ and the time step $k$ can be described as~\cite{heath2016overview}
\begin{align}
    &\boldsymbol{Y}^{p}_{\kappa,k} =\boldsymbol{H}^{p}_{\kappa,k} \boldsymbol{S}^{p}_{\kappa} + \boldsymbol{N}^{p}_{\kappa,k}, \label{eq:FreqObservation} 
\end{align}
with $\boldsymbol{S}^{p}_{\kappa}$ denoting the potentially pre-coded pilot signal over subcarrier $\kappa$, $\boldsymbol{Y}^{p}_{\kappa,k}$ denoting the received signal over subcarrier $\kappa$, $\boldsymbol{N}^{p}_{\kappa,k}$ denoting the additive white Gaussian noise, and $\boldsymbol{H}^{p}_{\kappa,k}$ being the channel frequency response. The transmitted signals may be bounced back to the $p$-th BS from multiple targets, and we assume that $\boldsymbol{H}^{p}_{\kappa,k}$ can be expressed as
\begin{align}
    &\boldsymbol{H}_{\kappa,k}^{p} = (\boldsymbol{W}_{k}^{p})^{\mathsf{H}}\sum _{i=1}^{I^{p}_{k}}g_{k}^{p,i}\boldsymbol{a}^{p}_{\text{BS}}(\boldsymbol{\theta}_{k}^{p,i})(\boldsymbol{a}^{p}_{\text{BS}}(\boldsymbol{\theta}_{k}^{p,i}))^{\mathsf{H}}e^{-\jmath 2\pi \kappa \Delta f \tau_{k}^{p,i}},  \notag 
\end{align}
where $\boldsymbol{W}_{k}^{p}$ represents a combining matrix at the $p$-th BS side, $\boldsymbol{a}^{p}_{\text{BS}}(\cdot)$ denotes the steering vector of the antenna array of the $p$-th BS, and $\Delta f$ denotes the subcarrier spacing. There are $I^{p}_{k}$ visible objects in the \ac{fov} of the $p$-th BS, which can bounce signals back to the corresponding BS and generates $I^{p}_{k}$ paths by assuming a point object model. Each path can be described as a complex gain $g_{k}^{p,i}$, a \ac{TOA} $\tau_{k}^{p,i}$, and an \ac{AOA} pair $\boldsymbol{\theta}_{k}^{p,i}$ in both azimuth and elevation. As the \ac{bs} can only receive bounced signals, there is no \ac{los} path, and  the \ac{bs} is always synchronized with itself, so there is no clock bias issue, resulting in $\tau_{k}^{p,i}=2\Vert \boldsymbol{s}_{\text{BS}}^{p}-\boldsymbol{s}_{k}^{i}\Vert/c$ for a target with 3D position $\boldsymbol{s}_{k}^{i}$, where $c$ denotes the speed of light. Moreover, we assume only single-bounce paths exist, therefore, the \ac{AOA} is equal to the corresponding \ac{AOD}. The relations between \ac{AOA}/\ac{AOD}  and the geometric state can be found in \cite[Sec.~2.2]{ge2024single}. Note that Doppler effects are not considered due to the short transmission interval. However, incorporating Doppler measurements would significantly enhance the detection and tracking performance of the mobile targets.
\vspace{-2mm}
\subsection{Measurement Models}\label{sec:GeoModel}
At each \ac{bs} side, a channel estimator  \cite{richter2005estimation,venugopal2017channel,jiang2021beamspace} is  applied on the received signals $\boldsymbol{Y}^{p}_{\kappa,k}$ in \eqref{eq:FreqObservation} to obtain the estimates of channel parameters of \ac{TOA} and \ac{AOA}, given the information on the sent signals $\boldsymbol{S}^{p}_{\kappa}$ and the combining matrix $\boldsymbol{W}_{k}^{p}$. To focus this work, we assume 
the channel parameter estimates are directly available to be used as measurements for sensing purposes. We model the measurements provided by the channel estimator at the $p$-th BS and the time step $k$ as a \ac{rfs}, i.e., $\mathcal{Z}_{k}^{p}=\{\boldsymbol{z}_{k}^{p,1},\dots, \boldsymbol{z}_{k}^{p,\hat{{I}}^{p}_{k}} \}$, with $\hat{{I}}^{p}_{k}$ denoting the number of resolved paths. Note that usually $\hat{{I}}^{p}_{k}\neq {I}^{p}_{k}$, as there could be also some clutter measurements originated from noise peaks during channel estimation, and not all paths can be resolved resulting in misdetections of some targets. These clutter measurements are modeled as a \ac{PPP}, parameterized by the intensity function $c(\boldsymbol{z})$. Since targets may also be undetected, we introduce the detection probability $p_{\text{D}}(\boldsymbol{s}^{p}_{\text{BS}},\boldsymbol{s}_{k}^{i}) \in [0,1]$ to capture the possibility of obtaining a measurement from an object located at $\boldsymbol{s}^{i}$, detected by the $p$-th BS. Assuming that the measurement $\boldsymbol{z}^{p,i}$ originates from the object located at $\boldsymbol{s}^{i}$, its likelihood is modeled as
    \begin{align}
 f(\boldsymbol{z}_{k}^{p,i}|\boldsymbol{s}^{p}_{\text{BS}},\boldsymbol{s}_{k}^{i})=\mathcal{N}(\boldsymbol{z}_{k}^{p,i};\boldsymbol{h}(\boldsymbol{s}^{p}_{\text{BS}},\boldsymbol{s}_{k}^{i}),\boldsymbol{R}_k^{p,i}),\label{pos_to_channelestimation}
\end{align}
where 
$\boldsymbol{h}(\boldsymbol{s}^{p}_{\text{BS}},\boldsymbol{s}_{k}^{i})=[\tau_{k}^{p,i},(\boldsymbol{\theta}_{k}^{p,i})^{\mathsf{T}}]^{\mathsf{T}}$ represents the nonlinear function that transforms the geometric information to \ac{TOA} and \ac{AOA} \cite[Sec.~2.2]{ge2024single}, and $\boldsymbol{R}_k^{p,i}$ denotes the known associated measurement covariance. However, it is important to note that the \ac{DA} remains unknown, making it unclear which source each measurement originates from.
\vspace{-2mm}

\section{Decentralized MTT per BS}
\vspace{-1mm}
In this section, we describe how each \ac{bs} tracks targets within its \ac{fov} over time using a \ac{TPMBM} filter \cite{garcia2020trajectory}, which estimates a set of target trajectories rather than a set of individual target states, enabling trajectory estimation without increasing the computational complexity. We will introduce the trajectory state representation and the high-level operation of the \ac{TPMBM} filter. For additional details on the the \ac{TPMBM} density, we refer to the Appendix. For notational brevity, we drop the BS index $p$ throughout this section.

\vspace{-2mm}
\subsection{Trajectory State Representation}
As described in Section~\ref{Sec:statemodels}, a single target state contains essential information about the target, such as its 3D position and 3D velocity. However, our objective is to track targets over time, aiming to estimate target trajectories rather than individual target states at specific time steps. A target trajectory is a finite sequence of target states, starting at any time step and ending at a later one. The trajectory state of a target can be modeled as a tuple \cite{svensson2014target,garcia2019multiple}
 $\boldsymbol{X}=(\mu,\nu,\boldsymbol{x}_{\mu:\nu})$, 
where $\mu$ denotes the birth time step of the trajectory, i.e, the time step when the trajectory begins, $\nu$ indicates the time step of the trajectory's most recent state, i.e, the time step when it ends, and $\boldsymbol{x}_{\mu:\nu}$ denotes the sequence of target states from the time step $\mu$ to the time step $\nu$, i.e., $\boldsymbol{x}_{\mu},\dots,\boldsymbol{x}_{\nu}$. If the current time step is $k$, then $\nu=k$ represents the trajectory is still ongoing, whereas $\nu<k$ signifies that the trajectory ended at time step $\nu$. The trajectory cannot end before it begins so that $ \mu \leq \nu$, and $(\nu-\mu+1)$ represents the length of the trajectory. Similar to a set of targets, we can model a set of trajectory as a \ac{rfs}  $\mathcal{X}=\{\boldsymbol{X}^{1},\dots,\boldsymbol{X}^{|\mathcal{X}|}\}$, modeled as a 
\ac{TPMBM} \ac{rfs}.

\subsection{TPMBM Filter}
Each \ac{bs} needs to track the trajectories of all targets $\mathcal{X}_{k}$ within its \ac{fov} over time, using the received measurements $\mathcal{Z}_{k}$ (as introduced in Section \ref{sec:GeoModel}) as input every time step. This can be accomplished by implementing a \ac{TPMBM} filter \cite{granstrom2018poisson,garcia2020trajectory} on each \ac{bs}. The primary objective of the \ac{TPMBM} filter is to recursively compute the posterior $f(\mathcal{X}_{k}|\mathcal{Z}_{k})$ at each time step, following the prediction and update steps of the Bayesian filtering framework for trajectory \acp{rfs} \cite{garcia2020trajectory}
\begin{align}
     f(\mathcal{X}_{k+1}|\mathcal{Z}_{1:k})& = \int   f(\mathcal{X}_{k}|\mathcal{Z}_{1:k})f(\mathcal{X}_{k+1}|\mathcal{X}_{k})\delta \mathcal{X}_{k}, \label{predict_Bay}\\
     f(\mathcal{X}_{k+1}|\mathcal{Z}_{1:k+1})& = \frac{f(\mathcal{X}_{k+1}|\mathcal{Z}_{1:k+1})\ell(\mathcal{Z}_{k+1}|\mathcal{X}_{k+1})}{\int   f(\mathcal{X}_{k+1}|\mathcal{Z}_{1:k})\ell(\mathcal{Z}_{k+1}|\mathcal{X}_{k+1})\delta \mathcal{X}_{k+1}},\label{update_Bay}
\end{align}
where $f(\mathcal{X}_{k+1}|\mathcal{X}_{k})$ denotes the transition density of the trajectory set, $\ell(\mathcal{Z}_{k+1}|\mathcal{X}_{k+1})$ denotes the RFS likelihood function of the measurement set give the trajectory set, and  $\int \psi(\mathcal{X})\delta \mathcal{X}$ indicates the trajectory set integral \cite{garcia2019multiple}. The expressions \eqref{predict_Bay}--\eqref{update_Bay} effectively serve to predict and update the TPMBM components. For further details, please refer to \cite{granstrom2018poisson,garcia2020trajectory}.

The output of the TPMBM filter after each time step $k$ is a  TPMBM density, which has the following components: $\lambda_{k|k}(\boldsymbol{X})$ represents the intensity function of the undetected  trajectories, while  $\{w_{k|k}^{j},\{r_{k|k}^{j,i},f_{k|k}^{j,i}(\boldsymbol{X})\}_{i\in \mathbb{I}_{k|k}^{j}}\}_{j\in \mathbb{I}_{k|k}}$ 
represents the state densities of the trajectories of the detected targets under different  \ac{DA} hypotheses.  Here, $w_{k|k}^{j}$  is a global hypothesis weight, $r_{k|k}^{j,i}$ is the existence probability of the $i$-th target under the $j$-th hypothesis, and $f_{k|k}^{j,i}(\boldsymbol{X})$ is the corresponding trajectory state density, which can be factorized into $f_{k|k}^{j,i}(\boldsymbol{x}_{\mu:\nu}|\mu,\nu)$ and $f_{k|k}^{j,i}(\mu,\nu)$ \cite[Eq.~4]{granstrom2018poisson}. In this paper, we focus on the Gaussian implementation of the TPMBM filter \cite[Sec.~V]{garcia2020trajectory}.

\section{MTT with Target Handover}
Each \ac{bs} locally applies a \ac{TPMBM} filter to track targets within its \ac{fov}. However, since each \ac{bs} has a limited \ac{fov}, and the targets move over time, some targets will inevitably exit the \acp{fov} of some \acp{bs} and enter the \acp{fov} of neighboring \acp{bs}. Although a \ac{bs} may lose track of targets as they leave its \acp{fov}, it can hand over these targets to the neighboring \acp{bs} to maintain continuous tracking.\footnote{We do not consider the fully centralized approach, which requires BSs to share all received measurements to a fusion center.} 
This section introduces the algorithm designed for efficient target handover between \acp{bs}.

\subsection{Handover Information}
We assume that all BSs can communicate with each other, and each BS is aware of its own and all other BSs' \acp{fov}. Target handover only occurs between neighboring BSs, as a target leaving one BS's \ac{fov} can only enter the \acp{fov} of nearby BSs. To reduce communication load, a BS only shares information about tracked target trajectories with the neighboring BSs into whose \acp{fov} those targets are expected to enter. This communication happens after the prediction step of the local \ac{TPMBM} filter at each BS. 

The first step is to determine if any targets are likely to move into another BS's \ac{fov} at each BS. Consider the $m$-th BS at time step $k+1$, suppose the predicted prior density $f(\mathcal{X}^{m}_{k+1}|\mathcal{Z}^{m}_{1:k})$ consists of the TPMBM components $\lambda^{m}_{k+1|k}(\boldsymbol{X})$ and $\{w_{k+1|k}^{m,j},\{r_{k+1|k}^{m,j,i},f_{k+1|k}^{m,j,i}(\boldsymbol{X})\}_{i\in \mathbb{I}_{k|k}^{m,j}}\}_{j\in \mathbb{I}_{k|k}^{m}}$. The $j,i$-th predicted trajectory is said to be entering the \ac{fov} of the $m'$-th BS, if 
\begin{equation}
    r_{k+1|k}^{m,j,i}\int  f_{k+1|k}^{m,j,i}(\boldsymbol{s}_{k+1}) p_{\text{D}}(\boldsymbol{s}_{k+1},\boldsymbol{s}^{m'}_{\text{BS}})\text{d}\boldsymbol{s}_{k+1} \geq \gamma,\label{fov}
\end{equation}
where $f_{k+1|k}^{m,j,i}(\boldsymbol{s}_{k+1})$ represents the predicted position density of the tracked target at the $m$-th BS at time step $k+1$, obtained during the prediction step of the TPMBM filter, $p_{\text{D}}(\boldsymbol{s}_{k+1},\boldsymbol{s}^{m'}_{\text{BS}})$ denotes the detection probability at the $m'$-th BS, and $\gamma$ is a predetermined threshold. Second, in case \eqref{fov} holds true, the $m$-th BS should inform the $m'$-th BS of the target’s trajectory details, facilitating the handover of tracking responsibilities. The trajectory information that needs to be shared includes the existence probability
$r_{k+1|k}^{m,j,i}$, the trajectory density $f_{k+1|k}^{m,j,i}(\boldsymbol{X})$, and the global hypothesis weight $w_{k+1|k}^{m,j}$. To efficiently hand over this information, we approximate it using a trajectory \ac{PPP},\footnote{We approximate the information as a trajectory PPP rather than directly adding a trajectory Bernoulli to the predicted prior density because the handed-over target has not yet been detected by the $m'$-th BS. Since its entry into the $m'$-th BS's \ac{fov} is based solely on our predictions, which may not materialize. Thus, it is more appropriate for the $m'$-th BS to treat this target trajectory as it would with other potentially detectable trajectories that have yet to be observed. Additionally, incorporating the trajectory PPP intensity allows the system to effectively manage repeated information exchanges between the two BSs when the target remains within both \acp{fov} for a duration. The TPMBM filter is more inclined to maintain an existing trajectory rather than initiate a new one from a previously undetected trajectory; therefore, only the trajectory from the initial handover will be retained.} and the trajectory \ac{PPP} intensity is formulated as follows
\begin{equation}
    \lambda_{\text{H},k+1|k}^{m,j,i}=w_{k+1|k}^{m,j}r_{k+1|k}^{m,j,i}f_{k+1|k}^{m,j,i}(\boldsymbol{X})\label{PPP_handover}.
\end{equation}
This PPP intensity can be seamlessly incorporated into the predicted trajectory PPP intensity of the $m'$-th BS $\lambda^{m'}_{k+1|k}(\boldsymbol{X})$, and the resulting combined trajectory PPP intensity is expressed as
\begin{equation}
    \lambda^{m'}_{k+1|k}(\boldsymbol{X})\gets \lambda^{m'}_{k+1|k}(\boldsymbol{X})+\lambda_{\text{H},k+1|k}^{m,j,i}(\boldsymbol{X})\label{PPP_handover_new}.
\end{equation}
{To minimize unnecessary handovers and corresponding inter-BS data exchange, each BS only hands over targets that have not been recently shared with the destination BS.}\footnote{Full details of the management of list of recently handed over target and hyper-parameters to avoid ping-ponging of targets are omitted for space reasons. }
\vspace{-3mm}
\subsection{Target Handover Algorithm}
We examine all the detected trajectories at each BS to identify any trajectories that may enter the \acp{fov} of neighboring BSs every time step, adding the corresponding trajectory PPP intensities to the predicted trajectory PPP intensities. Meanwhile, the trajectory MBM components remain unchanged, as no modifications occur in that regard. The resulting algorithm is summarized in Algorithm \ref{alg:handover}.
\vspace{-3mm}
\begin{algorithm}
\caption{Target Handover Algorithm} \label{alg:handover}
\begin{algorithmic}[1]
\Require \parbox[t]{\dimexpr\linewidth- \algorithmicindent * 1}{ Predicted  $\{w_{k+1|k}^{m,j},\{r_{k+1|k}^{m,j,i},f_{k+1|k}^{m,j,i}(\boldsymbol{X})\}_{i\in \mathbb{I}_{k|k}^{m,j}}\}_{j\in \mathbb{I}_{k|k}^{m}},\\\lambda^{m}_{k+1|k}(\boldsymbol{X}),
\forall  m \in\{1,\cdots,M\}$;
\strut}
\Ensure Reconstructed  $\lambda^{m}_{k+1|k}(\boldsymbol{X}),\forall m \in\{1,\cdots,M\}$;
\For{$m\in\{1,\cdots,M\}$, $j\in\mathbb{I}_{k|k}^{m}$, $i\in\mathbb{I}_{k|k}^{m,j}$, $m'\neq m$} 
\If{\eqref{fov} {and (not recently sent)}}
\State {Compute the handed-over PPP \eqref{PPP_handover};}
\State Communicate PPP to BS $m'$;
\State {Add \eqref{PPP_handover} to PPP  at BS $m'$ \eqref{PPP_handover_new};}
\EndIf
\EndFor
\end{algorithmic}
\end{algorithm}
\vspace{-3mm}

After reviewing all the trajectories and incorporating the resulting trajectory PPP intensities into the corresponding predicted trajectory PPP intensities, we update the density at each BS using the relevant measurement set, expressed as
\begin{align}
f(\mathcal{X}_{k+1}^{m}|\mathcal{Z}^{m}_{1:k+1})& = \frac{\tilde{f}(\mathcal{X}^{m}_{k+1}|\mathcal{Z}^{m}_{1:k+1})\ell(\mathcal{Z}^{m}_{k+1}|\mathcal{X}^{m}_{k+1})}{\int   \tilde{f}(\mathcal{X}^{m}_{k+1}|\mathcal{Z}^{m}_{1:k})\ell(\mathcal{Z}^{m}_{k+1}|\mathcal{X}^{m}_{k+1})\delta \mathcal{X}^{m}_{k+1}}
\end{align}
with $\tilde{f}(\mathcal{X}^{m}_{k+1}|\mathcal{Z}^{m}_{1:k+1})$ denoting the density for the $m$-th BS, as obtained from Algorithm \ref{alg:handover}, which involves adding the corresponding trajectory PPP intensities to its existing trajectory PPP intensity.

\vspace{-4mm}

\section{Numerical Results}
\vspace{-2mm}

\subsection{Simulation Environment}
\begin{figure}
\centerline{\includegraphics[width=0.8\linewidth]{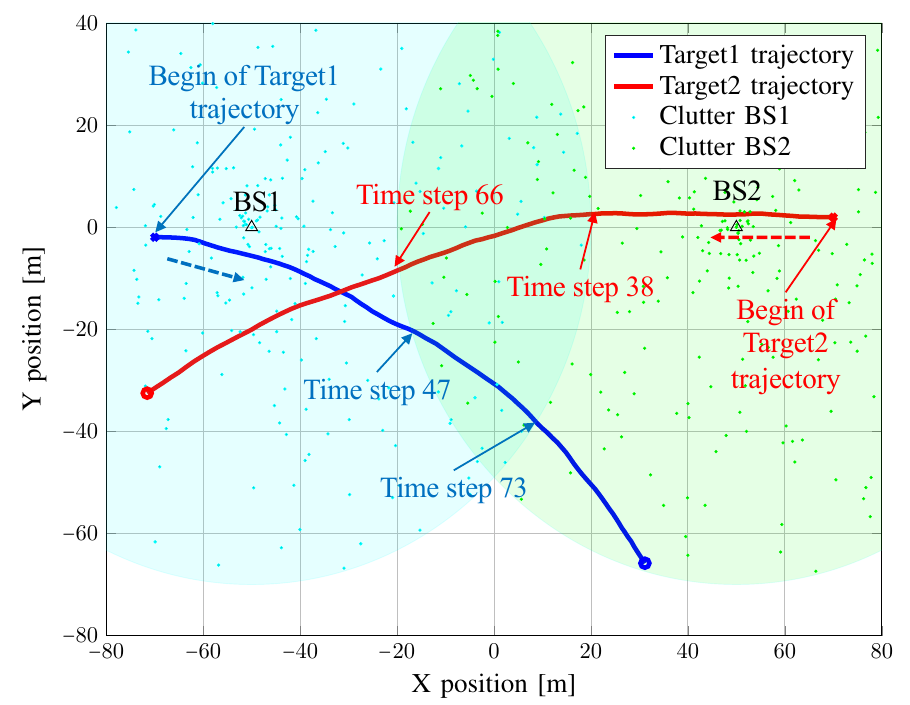}}
    \vspace{-2mm}
\caption{Simulated scenario with two BSs, and two targets. The clutter for all time step is also shown. Each BS has a limited \ac{fov}, while both targets follow a constant velocity movement. Target1 enters BS2's \ac{fov} at time step 47 and exits BS1's \ac{fov} at time step 73. Targets2 enters  BS1's \ac{fov} at time step 38 and exits  BS2's \ac{fov} at time step 66.}
\label{Fig.scenario}
\end{figure}

We consider a scenario depicted in Fig.~\ref{Fig.scenario}, where two BSs are positioned at $[\pm 50 \, \text{m},0 \, \text{m},10\, \text{m}]^{\mathsf{T}} $, respectively, and two targets follow a nearly constant velocity movement over time, with the acceleration noise standard deviation of $0.05\, \text{m/s}^{2}$ along both the $x$ and $y$ axes, and $0\, \text{m/s}^{2}$ along the $z$ axis. The sampling time is set to $100\, \text{ms}$, and 100 time steps are considered. Target1 starts at $[-70 \, \text{m},-2 \, \text{m},1.5\, \text{m}]^{\mathsf{T}} $ with an initial velocity $[14 \, \text{m/s},0 \, \text{m/s},0\, \text{m/s}]^{\mathsf{T}} $, while the target2 starts at $[70 \, \text{m},2 \, \text{m},1.5\, \text{m}]^{\mathsf{T}} $ with an initial velocity $[-14 \, \text{m/s},0 \, \text{m/s},0\, \text{m/s}]^{\mathsf{T}} $. The targets are only visible when they are in the \ac{fov} of the \acp{bs}, set to $70 \, \text{m}$. The detection probability is set to $0.9$ within the \ac{fov} of each BS and 0 outside it, the survival probability is set to $0.99$ for both BSs, and the threshold $\gamma$ is set to 0.5. The measurement noise covariance matrix is fixed to $\boldsymbol{R}=\text{diag}[ 0.1^{2} \, \text{m}^{2},0.01^{2} \, \text{rad}^{2},0.01^{2} \, \text{rad}^{2}]$. The clutter intensity $c(\boldsymbol{z})$ is set to $3U_{\text{FOV}}$ for both BSs, with $U_{\text{FOV}}$ representing a uniform distribution inside the \ac{fov} of each BS and $3$ representing the expected number of clutter measurements per time step. For the TPMBM implementation, the filters are configured with the following parameters \cite{garcia2020metric,garcia2020trajectory}: maximum number of
hypotheses is set to 200, the pruning threshold for PPP weights is set at $10^{-5}$, the pruning threshold for  Bernoulli components is set at $10^{-5}$, and a $5$-scan implementation is employed. The benchmark algorithm involves independent operation of the BSs, each running a TMPMB filter. The sensing performance is numerically evaluated using the \ac{RMS} trajectory error \cite{garcia2020trajectory} for the trajectory set and the \ac{RMSE} across the given trajectory. We conduct 100 Monte Carlo (MC) simulations for each algorithm, with the final results computed as the average across all independent simulations.

\subsection{Results and Discussion}
Fig.~\ref{Fig.RESULTSRMSEBS1} illustrates the \ac{RMS} trajectory error at BS1 over time for cases with and without the integration of the target handover method. The errors are identical in both scenarios before time step 38, as only target1 is within BS1’s \ac{fov}, resulting in the same tracking performance. After time step 38, target2 enters BS1's \ac{fov} and begins to be tracked, reducing the RMS trajectory error as shown by the decreasing blue dashed line. However, when the target handover scheme is not implemented, BS1 lacks access to target2’s earlier trajectory, causing the \ac{RMS} trajectory error to remain relatively high because part of the trajectory is missed. In contrast, when the target handover scheme is applied, BS1 inherits target2’s prior trajectory from BS2 and continues tracking it seamlessly, leading to a significant drop in the red solid line at time step 38. This results in a lower \ac{RMS} error that persists until time step 73. After time step 73, the errors for both cases increase when target1 leaves BS1’s \ac{fov}, leading to a loss of tracking. 

Fig.~\ref{Fig.RESULTSRMSETARGET2} displays the \ac{RMSE} of target2's trajectory at BS1 for both cases. The TPMBM filter effectively tracks target2's full trajectory when the target handover method is used, as BS1 inherits the trajectory information from BS2 at time step 38. Without the target handover scheme, BS1 can only track target2 after it enters its \ac{fov}, which is why the red solid line starts at the beginning while the blue dashed line starts at time step 38. Notably, the target handover mechanism provides more accurate estimates immediately after target2 enters BS1's \ac{fov}, as the red solid line stays below the blue dashed line between time steps 38 and 47. This is because BS1 inherits a lower-error and lower-covariance trajectory from BS2, allowing it to continue tracking target2 with improved accuracy and lower covariance, compared to starting from an untracked trajectory. However, this gap gradually narrows over time and eventually converges, as both cases receive the same measurements, reducing the impact of the initial error and covariance at time step 38.

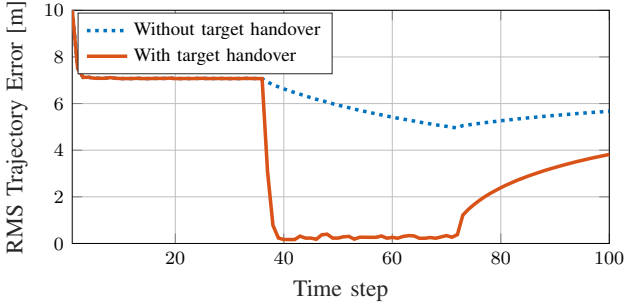
\begin{figure}
\center
%
%
\definecolor{mycolor1}{rgb}{0.00000,0.44700,0.74100}%
\definecolor{mycolor2}{rgb}{0.85000,0.32500,0.09800}%
\definecolor{mycolor3}{rgb}{0.49400,0.18400,0.55600}%
\definecolor{mycolor4}{rgb}{0.46600,0.67400,0.18800}%
\definecolor{mycolor5}{rgb}{0.30100,0.74500,0.93300}%
\begin{tikzpicture}[scale=1\columnwidth/10cm,font=\footnotesize]
\begin{axis}[%
width=8cm,
height=3.5cm,
at={(0in,0in)},
scale only axis,
unbounded coords=jump,
xmin=1,
xmax=100,
xlabel style={font=\color{white!15!black}},
xlabel={Time step},
ymin=0,
ymax=10,
ylabel style={font=\color{white!15!black}},
ylabel={RMS Trajectory Error [m]},
yminorticks=true,
xmajorgrids,
ymajorgrids,
legend style={at={(0.01,0.99)},  anchor=north west, legend cell align=left, align=left, draw=white!15!black}
]
\addplot [color=mycolor1, dotted, line width=1.5pt]
  table[row sep=crcr]{%
1	10\\
2	7.5101296004013\\
3	7.1226603885732\\
4	7.12853383438651\\
5	7.08814827709641\\
6	7.08495288198693\\
7	7.08320825442225\\
8	7.11267066945367\\
9	7.08469603766049\\
10	7.0726589009724\\
11	7.07257559561768\\
12	7.07247863174341\\
13	7.08324648667837\\
14	7.0723042310587\\
15	7.07222239161994\\
16	7.07658539809006\\
17	7.08043875684729\\
18	7.07795486983731\\
19	7.09060569555755\\
20	7.07199337031891\\
21	7.07195876690872\\
22	7.07514580097496\\
23	7.07192491729921\\
24	7.07779090811349\\
25	7.08318805225797\\
26	7.07188378792815\\
27	7.07449239782042\\
28	7.07565406284828\\
29	7.07428737833186\\
30	7.07184419965189\\
31	7.07410341457756\\
32	7.07182189783636\\
33	7.08039420239483\\
34	7.07495009087609\\
35	7.07385172498432\\
36	7.06855305397808\\
37	6.92073392049859\\
38	6.81188842428457\\
39	6.71703320422786\\
40	6.63061917342882\\
41	6.54935848425655\\
42	6.47094935113244\\
43	6.40087310539782\\
44	6.32419227333116\\
45	6.2536364431012\\
46	6.18374837677079\\
47	6.12641076926736\\
48	6.06402849676592\\
49	5.99339279943822\\
50	5.93323704668939\\
51	5.87735080576373\\
52	5.82146675663858\\
53	5.76150715287733\\
54	5.71122911585698\\
55	5.65912259688315\\
56	5.60836849973313\\
57	5.5590285383989\\
58	5.51409819232545\\
59	5.46258108999234\\
60	5.41844725479296\\
61	5.37236333680493\\
62	5.33270237077851\\
63	5.29250450562473\\
64	5.25029699185263\\
65	5.20466112542442\\
66	5.16518309156997\\
67	5.12801554333983\\
68	5.08878237143601\\
69	5.05473944770602\\
70	5.022141666047\\
71	4.98246106950141\\
72	4.95509980699454\\
73	5.05527264564375\\
74	5.08651160681276\\
75	5.11675186172529\\
76	5.14730386726663\\
77	5.17813698607494\\
78	5.20676608340511\\
79	5.23331657370213\\
80	5.2614556124974\\
81	5.28874747133947\\
82	5.31296986085854\\
83	5.33874327516762\\
84	5.36155822178801\\
85	5.3853761301845\\
86	5.40747505086941\\
87	5.42843287254436\\
88	5.45196434210505\\
89	5.47180026602012\\
90	5.49112799926902\\
91	5.51096634608495\\
92	5.53129126382695\\
93	5.54915905758702\\
94	5.57042634857988\\
95	5.58552199034625\\
96	5.60489927699796\\
97	5.62197660349149\\
98	5.63685844230438\\
99	5.6531739436613\\
100	5.6700006997449\\
};
\addlegendentry{Without target handover}

\addplot [color=mycolor2, line width=1.5pt]
  table[row sep=crcr]{%
1	10\\
2	7.5101296004013\\
3	7.1226603885732\\
4	7.12853383438651\\
5	7.08814827709641\\
6	7.08495288198693\\
7	7.08320825442225\\
8	7.11267066945367\\
9	7.08469603766049\\
10	7.0726589009724\\
11	7.07257559561768\\
12	7.07247863174341\\
13	7.08324648667837\\
14	7.0723042310587\\
15	7.07222239161994\\
16	7.07658539809006\\
17	7.08043875684729\\
18	7.07795486983731\\
19	7.09060569555755\\
20	7.07199337031891\\
21	7.07195876690872\\
22	7.07514580097496\\
23	7.07192491729921\\
24	7.07779090811349\\
25	7.08318805225797\\
26	7.07188378792815\\
27	7.07449239782042\\
28	7.07565406284828\\
29	7.07428737833186\\
30	7.07184419965189\\
31	7.07410341457756\\
32	7.07182189783636\\
33	7.08039420239483\\
34	7.07495009087609\\
35	7.07385172498432\\
36	7.06855305397808\\
37	3.09035004977371\\
38	0.778801419171165\\
39	0.233631583770248\\
40	0.168844092800225\\
41	0.168962374272091\\
42	0.1677659971248\\
43	0.314314274688921\\
44	0.229372272994269\\
45	0.230146082736852\\
46	0.180662887182406\\
47	0.37342003611637\\
48	0.397337007305832\\
49	0.230728184233817\\
50	0.230396101883444\\
51	0.28669619063214\\
52	0.301870953747282\\
53	0.185306651004003\\
54	0.267608285186827\\
55	0.266279122160297\\
56	0.263920876151696\\
57	0.263142419344247\\
58	0.321474116629432\\
59	0.226824838732042\\
60	0.260205880709128\\
61	0.22498746292839\\
62	0.301246140164795\\
63	0.337031093839291\\
64	0.323257038995742\\
65	0.223279054247293\\
66	0.223939879054347\\
67	0.25499703385935\\
68	0.223480571847264\\
69	0.281433087592582\\
70	0.338518343246227\\
71	0.266889958043038\\
72	0.378321675358903\\
73	1.21666623925404\\
74	1.45685970395622\\
75	1.65760618879449\\
76	1.83564189672448\\
77	1.99706620296569\\
78	2.13967517109496\\
79	2.26729286019475\\
80	2.39040101939258\\
81	2.50463024261743\\
82	2.60664101840233\\
83	2.70690564848828\\
84	2.79706032189721\\
85	2.88542302982966\\
86	2.96725498076749\\
87	3.04410870649102\\
88	3.12284959881529\\
89	3.19269071338873\\
90	3.25953635619933\\
91	3.32527187342896\\
92	3.38995477387377\\
93	3.4489403245404\\
94	3.51179656701127\\
95	3.56341565992691\\
96	3.62042549968371\\
97	3.67259337007038\\
98	3.72026594593787\\
99	3.76904904259697\\
100	3.81756328471914\\
};
\addlegendentry{With target handover}

\end{axis}
\end{tikzpicture}%
\vspace{-2mm}
\caption{Comparison of the overall tracking performances of the TPMBM filter with and without the integration of the target handover method at BS1.}
\label{Fig.RESULTSRMSEBS1}
\end{figure}
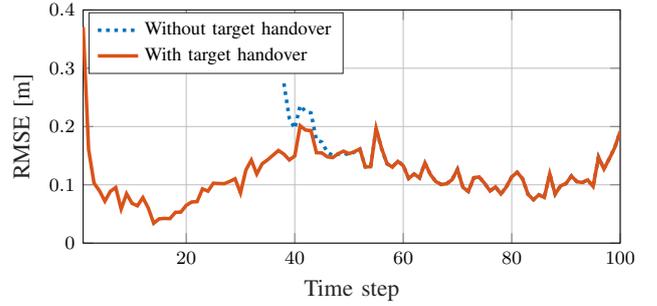
\begin{figure}
\center
%
%
\definecolor{mycolor1}{rgb}{0.00000,0.44700,0.74100}%
\definecolor{mycolor2}{rgb}{0.85000,0.32500,0.09800}%
\definecolor{mycolor3}{rgb}{0.49400,0.18400,0.55600}%
\definecolor{mycolor4}{rgb}{0.46600,0.67400,0.18800}%
\definecolor{mycolor5}{rgb}{0.30100,0.74500,0.93300}%
\begin{tikzpicture}[scale=1\columnwidth/10cm,font=\footnotesize]
\begin{axis}[%
width=8cm,
height=3.5cm,
at={(0in,0in)},
scale only axis,
unbounded coords=jump,
xmin=1,
xmax=100,
xlabel style={font=\color{white!15!black}},
xlabel={Time step},
ymin=0,
ymax=0.4,
ylabel style={font=\color{white!15!black}},
ylabel={RMSE [m]},
yminorticks=true,
xmajorgrids,
ymajorgrids,
legend style={at={(0.01,0.99)},  anchor=north west, legend cell align=left, align=left, draw=white!15!black}
]
\addplot [color=mycolor1, dotted, line width=1.5pt]
  table[row sep=crcr]{%
1	inf\\
2	inf\\
3	inf\\
4	inf\\
5	inf\\
6	inf\\
7	inf\\
8	inf\\
9	inf\\
10	inf\\
11	inf\\
12	inf\\
13	inf\\
14	inf\\
15	inf\\
16	inf\\
17	inf\\
18	inf\\
19	inf\\
20	inf\\
21	inf\\
22	inf\\
23	inf\\
24	inf\\
25	inf\\
26	inf\\
27	inf\\
28	inf\\
29	inf\\
30	inf\\
31	inf\\
32	inf\\
33	inf\\
34	inf\\
35	inf\\
36	inf\\
37	inf\\
38	0.273976920344954\\
39	0.214187831649233\\
40	0.196562122166298\\
41	0.236081588272293\\
42	0.225633441836874\\
43	0.223251575639288\\
44	0.176471375388968\\
45	0.173312040774422\\
46	0.155818511228141\\
47	0.150442457477752\\
48	0.152532246252908\\
49	0.154248183873937\\
50	0.151488081423803\\
51	0.156271386842209\\
52	0.160923404014955\\
53	0.130923138002384\\
54	0.131059898980175\\
55	0.196215125972201\\
56	0.162094675712772\\
57	0.135921547727292\\
58	0.130738585856314\\
59	0.14074400879993\\
60	0.133251038100031\\
61	0.111054361515838\\
62	0.118478879820381\\
63	0.111772250736947\\
64	0.13753659073732\\
65	0.117768201302617\\
66	0.10563645446489\\
67	0.100539602804712\\
68	0.101886333314997\\
69	0.108747795925894\\
70	0.126495747945386\\
71	0.0956891818806059\\
72	0.0887224457203826\\
73	0.112016677505341\\
74	0.113544749618313\\
75	0.10248176007498\\
76	0.0894396585438601\\
77	0.0964542940702853\\
78	0.084386026320464\\
79	0.0963470244858429\\
80	0.11363064144048\\
81	0.121473192099575\\
82	0.110306922328999\\
83	0.0837982243808009\\
84	0.074439616124305\\
85	0.0828746705058796\\
86	0.0783534667189759\\
87	0.118775102672695\\
88	0.0839535426772426\\
89	0.0982555040758475\\
90	0.102106210942097\\
91	0.115085417586634\\
92	0.105455256088207\\
93	0.103954679183898\\
94	0.108378645596485\\
95	0.0972013929072437\\
96	0.147331830282938\\
97	0.1270689506365\\
98	0.144999173669613\\
99	0.164086615333408\\
100	0.191598035219746\\
};
\addlegendentry{Without target handover}

\addplot [color=mycolor2, line width=1.5pt]
  table[row sep=crcr]{%
1	0.370766007383987\\
2	0.16063526525548\\
3	0.102890259452236\\
4	0.0897435102694914\\
5	0.0719675692626421\\
6	0.0887104314632595\\
7	0.0953595310210106\\
8	0.0585414141356765\\
9	0.0847338250547318\\
10	0.0682655596093035\\
11	0.0640079131893214\\
12	0.0780365592411265\\
13	0.0602876520118942\\
14	0.0339463493372735\\
15	0.041612105192783\\
16	0.042317468935593\\
17	0.0420354988260133\\
18	0.0528330580784574\\
19	0.0531549692985239\\
20	0.0647927599489949\\
21	0.0705475435992835\\
22	0.0711046232219275\\
23	0.0930291559954502\\
24	0.0891036602583141\\
25	0.102768803519019\\
26	0.102260305272364\\
27	0.10192007625145\\
28	0.105727106911743\\
29	0.110303404047742\\
30	0.0861927961750512\\
31	0.125417199298324\\
32	0.14240415593544\\
33	0.118485118060056\\
34	0.136148464859824\\
35	0.14283687446549\\
36	0.150912402865341\\
37	0.158888843525593\\
38	0.152623720711781\\
39	0.142647472525968\\
40	0.149340262323319\\
41	0.201364731675826\\
42	0.19390330243201\\
43	0.193026183071767\\
44	0.155080567631793\\
45	0.154681266956571\\
46	0.148030635406356\\
47	0.146683237176353\\
48	0.153268287061134\\
49	0.157774931394923\\
50	0.153361739189234\\
51	0.156873020861885\\
52	0.1615545016795\\
53	0.130638869532618\\
54	0.131045565699149\\
55	0.197100312258237\\
56	0.162976846243686\\
57	0.136584528534054\\
58	0.130253616859621\\
59	0.140191271345068\\
60	0.13291046018921\\
61	0.110910026620397\\
62	0.118536628779577\\
63	0.111817155004601\\
64	0.137563464243439\\
65	0.117768098858807\\
66	0.105648086214466\\
67	0.100543115698378\\
68	0.101895897280822\\
69	0.108759155432859\\
70	0.126508068046763\\
71	0.0956875732582307\\
72	0.0887213776745813\\
73	0.112015685352096\\
74	0.113545080361322\\
75	0.10248244830898\\
76	0.0894399361839105\\
77	0.0964537092045905\\
78	0.0843859990434108\\
79	0.0963472100367262\\
80	0.113630776765656\\
81	0.121473212940186\\
82	0.110306905288573\\
83	0.0837982264582268\\
84	0.0744396122612024\\
85	0.0828746555336997\\
86	0.0783534702798794\\
87	0.11877510021081\\
88	0.0839535408048188\\
89	0.09825550429496\\
90	0.102106210640501\\
91	0.115085417539906\\
92	0.105455256091229\\
93	0.103954678943084\\
94	0.108378645464684\\
95	0.0972013930228152\\
96	0.147331830585337\\
97	0.127068950753216\\
98	0.144999173713005\\
99	0.164086615293795\\
100	0.191598035174912\\
};
\addlegendentry{With target handover}

\end{axis}
\end{tikzpicture}%
\vspace{-2mm}
\caption{Comparison of the tracking performances of the TPMBM filter with and without the integration of the target handover method for target2 at BS1.}
\label{Fig.RESULTSRMSETARGET2}
\end{figure}

\begin{figure}
\centerline{\includegraphics[width=0.85\linewidth]{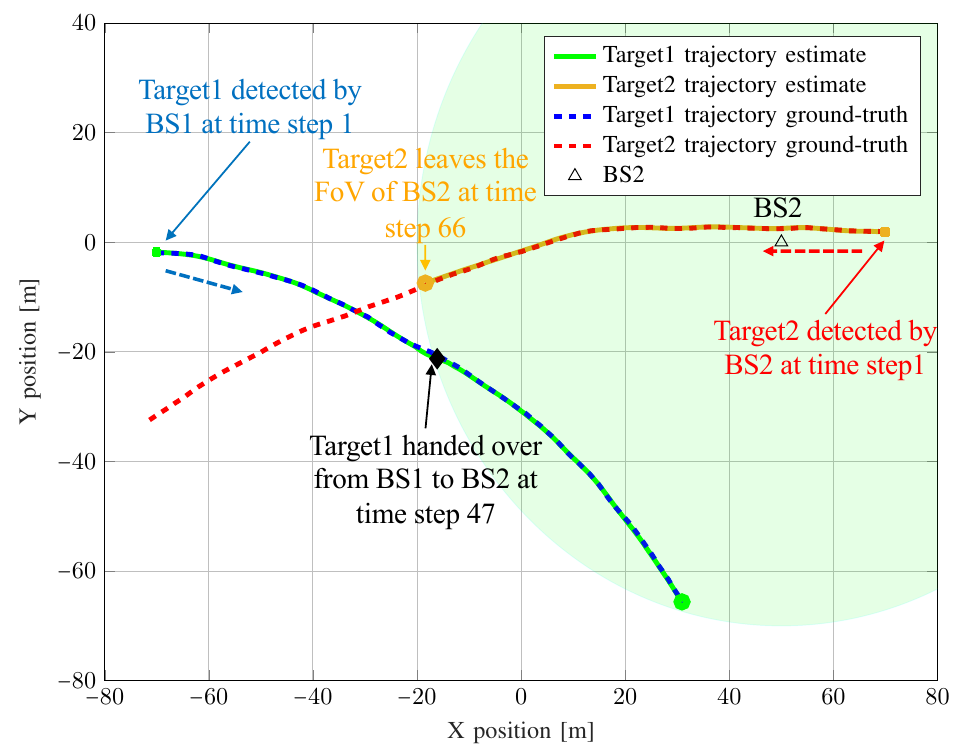}}
    \vspace{-2mm}
\caption{Tracking results from the TPMBM filter that integrates the target handover method at BS2. {Clutter is present but not shown in this visualization.}}
\label{Fig.RESULTSBS2}
\end{figure}

Fig.~\ref{Fig.RESULTSBS2} shows the trajectory tracking results from the TPMBM filter at BS2, with the integration of the target handover method. The results show that the TPMBM filter at BS2, when combined with the target handover method, successfully tracks a portion of the target2's trajectory, as indicated by the alignment of the yellow solid line with a section of the red dashed line. Additionally, it tracks the full trajectory of target1, from its origin to the end, indicated by the alignment of the green solid line with the blue dashed lines. Although BS2 cannot track target1 prior to time step 47, since it had not yet entered the BS2's \ac{fov}, BS1 hands over all trajectory information about target1 to BS2 at the prediction step of time step 47, just as the target is about to enter the BS2's \ac{fov}. BS2 then continues to track target1 based on the inherited trajectory information, resulting in an accurate estimation of target1's full trajectory. Nevertheless, BS2 cannot continue tracking target2 once it leaves its \ac{fov} at time step 66. As target2 moved into BS1’s \ac{fov}, BS2 handed over tracking responsibility to BS1. However, once target2 exits BS2's \ac{fov}, the trajectory information is not relayed back to BS2, resulting in a temporary loss of tracking.  Notably, during time steps 38 to 66, both BSs track target2 independently, as it remains within the \acp{fov} of BSs.

\vspace{-1mm}
\section{Conclusions}
In this paper, we address the target handover problem in DISAC systems by introducing a novel RFS-based target handover algorithm. We provide detailed execution guidelines, including what information to share, when, where, and how it should be transmitted during handover. The proposed target handover mechanism is integrated with the TPMBM filter to address the sensing challenges at each BS. Through simulations, we show that the proposed method efficiently facilitates target handover between BSs as targets enter their \acp{fov}, enabling each BS to track the complete trajectories of targets within its \ac{fov} from initial detection until exit. The results also demonstrate that the target handover improves the TPMBM filter’s sensing performance, particularly for targets that have recently entered a BS's \ac{fov}.
\vspace{-2mm}
\appendix[Basics of TPMBM Density]
The structure of the \ac{TPMBM} \ac{rfs} resembles that of the \ac{PMBM} target \ac{rfs}. In this framework, $\mathcal{X}$ is viewed as the union of two disjoint trajectory sets $\mathcal{X}_{\mathrm{U}}$ and $\mathcal{X}_{\mathrm{D}}$. Here,  $\mathcal{X}_{\mathrm{U}}$ is the set of undetected trajectories that are hypothesized to exist but never have been detected, i.e., no measurement has been associated with these trajectories. Conversely, $\mathcal{X}_{\mathrm{D}}$ denotes the set of detected trajectories that exist and have been detected at least once, i.e, associated with at least one measurement before \cite{garcia2018poisson,garcia2019multiple}. We model $\mathcal{X}_{\mathrm{U}}$ as a trajectory \ac{PPP}, with the density described by 
\begin{equation}
    f_{\mathrm{P}}(\mathcal{X}_{\mathrm{U}})=e^{-\int\lambda(\boldsymbol{X}')\mathrm{d}\boldsymbol{X}'}\prod_{\boldsymbol{X} \in \mathcal{X}_{\mathrm{U}} }\lambda(\boldsymbol{X}),\label{PPP}
\end{equation}
where $\lambda(\cdot)$ represents the trajectory PPP intensity function, defined across the entire trajectory state space. This means that the realizations of the PPP are trajectories with a birth time, a time of the most recent state, and a sequence of states. We model  $\mathcal{X}_{\mathrm{D}}$ as a trajectory \ac{MBM}, with the density described by
\begin{equation}
    f_{\mathrm{MBM}}(\mathcal{X}_{\mathrm{D}})= \sum_{j \in \mathbb{I}}w^{j}\sum_{\mathcal{X}^{1}\biguplus \dots \biguplus \mathcal{X}^{n}=\mathcal{X}_{\mathrm{D}}}\prod_{i=1}^{|\mathcal{X}_{\mathrm{D}}|}f^{j,i}_{\mathrm{B}}(\mathcal{X}^{i}),\label{MBM}
\end{equation}
where  $\mathbb{I}$ denotes the index set of all global hypotheses, and $w^{j}\ge 0$ represents the weight associated with the $j$-th global hypothesis, satisfying $\sum_{j\in\mathbb{I}}w^{j}=1$ \cite{williams2015marginal}. The term $f_{\mathrm{B}}^{j,i}(\cdot)$ refers to the trajectory Bernoulli density of the $i$-th trajectory under the $j$-th global hypothesis. Similar to the target Bernoulli, each trajectory Bernoulli follows
\begin{equation}
f^{j,i}_{\mathrm{B}}(\mathcal{X}^{i})=
\begin{cases}
1-r^{j,i} \quad& \mathcal{X}^{i}=\emptyset \\ r^{j,i}f^{j,i}(\boldsymbol{X}) \quad & \mathcal{X}^{i}=\{\boldsymbol{X}\} \\ 0 \quad & \mathrm{otherwise}
\end{cases}
\end{equation} 
where $r^{j,i} \in [0,1]$ denotes the existence probability, and $f^{j,i}(\cdot)$ represents the trajectory density. For more information on the trajectory PPP and MBM densities, please refer to \cite{garcia2018poisson,garcia2019multiple}.
The density of $\mathcal{X}$ can then be calculated using the convolution formula \cite[eq. (4.17)]{mahler2014advances} as
\begin{equation}
    f(\mathcal{X})=\sum_{\mathcal{X}_{\mathrm{U}}\biguplus\mathcal{X}_{\mathrm{D}}=\mathcal{X}}f_{\mathrm{P}}(\mathcal{X}_{\mathrm{U}})f_{\mathrm{MBM}}(\mathcal{X}_{\mathrm{D}}),\label{PMBM}
\end{equation}
which can also be parameterized by $\lambda(\boldsymbol{X})$ and $\{w^{j},\{r^{j,i},f^{j,i}(\boldsymbol{X})\}_{i\in \mathbb{I}^{j}}\}_{j\in \mathbb{I}}$, with $\mathbb{I}^{j}$ representing the index set of trajectories (i.e., the Bernoulli components) under the $j$-th global hypothesis.  

\balance
\bibliography{IEEEabrv,Bibliography}

\end{document}